\documentclass[
    ,final            
  ]
  {aipproc}

\layoutstyle{6x9}

\begin{document}
\title{Statistics of turbulence via polarimetry: alignment of grains and atoms}


\author{A. Lazarian \& Huirong Yan}{address={Department of Astronomy, 
University of Wisconsin, 475 N. Charter
St., Madison, WI 53706}}

\begin{abstract}
Most astrophysical fluids are turbulent and magnetized. Fluctuations
of polarization provide a promising way to study
astrophysical magnetic turbulence. We discuss 
polarization that arises from grains and atoms  aligned in respect
to magnetic field, describe the processes of alignment, explain
when the alignment reflects the direction of magnetic field.
We show that temporal
fluctuations of interplanetary magnetic field can be studied
using aligned sodium atoms.

\end{abstract}
\maketitle

\section{I. Why do we care about astrophysical turbulence?}

It is generally accepted that most astrophysical fluids are turbulent 
with magnetic fields playing important role. In this situation the 
properties of the fluids depend
of the properties of the magnetized turbulence. In most cases astrophysical 
scales are much larger than the  scales determined by the plasma effects 
and therefore the Magneto-hydrodynamic ({MHD}) description of astrophysical 
fluids is adequate.

Scattering and acceleration of cosmic rays (see \cite{Sch99} and 
references therein), acceleration of charged interstellar dust particles 
\cite{LY02, YL03}, conduction of heat 
(see \cite{Choet, Naray}) depend intimately on
 the properties of MHD turbulence.  Rates at  which the topology of
otherwise ``frozen in'' astrophysical magnetic field changes, i.e. 
magnetic reconnection rates, according to
\cite{LV99} are controlled
by MHD turbulence\footnote{Note, that magnetic reconnection is essential 
for a wide variety of astrophysical processes from Solar flares 
to dynamo generation of magnetic field.}.

 It happened that for years MHD turbulence was a 
rather speculative area of research, where many models 
could coexist without much chance of being tested. 
An important progress in the field was stimulated by {\it in situ} 
measurements of the statistics of interplanetary  turbulence 
(see \cite{Mont, Hig}). However, these measurements
were not sufficiently detailed to get a good insight into 
the structure of the magnetized eddies. 

A study of 
{\it incompressible} MHD turbulence by Goldreich \& Sridhar
\cite{GS95} became 
a milestone for the subject. There a universal scaling was proposed for 
magnetic eddies in the system of reference related to the {\it local} 
mean magnetic field of the eddy. Considerations applied by
 Goldreich \& Sridhar to pseudo-Alfv\'en waves have been generalized and 
extended to slow modes  of {\it compressible} MHD 
(\cite{LG01, CL02a}).  
Recent theoretical and numerical work (see \cite{CL03}, \cite{CL04, CL05} 
\cite{CLV, LVC}) shed 
light on what is happening with MHD turbulence in partially ionized gas, 
on turbulence intermittency, on coupling of compressible and Alfv\'enic modes, 
on rates of turbulence decay etc. 
In its turn this work stimulated revisions of 
some earlier paradigms for basic astrophysical processes 
(see \cite{YL02, YL03, Choet, YL04}).

It is important to stress that for the first time ever the theoretical 
predictions were tested with direct 3D numerical simulations. Does this mean
that we are sure that we know the final truth about the
actual astrophysical turbulence? Surely, not. It is
well known that  flows are similar when their Reynolds numbers, which
are the ratio of the advection to dissipation terms in the Navie-Stocks
equations are the same. For MHD, magnetic diffusivity is important
and therefore magnetic Reynolds numbers should be also the
same. As the corresponding numbers for the numerical simulations and
astrophysics are different by a factor  of $10^6$ and larger, from this 
point of view, the flows simulated in computers and those in
astrophysical fluids are very different. 

In other words, while theoretical expectations from elegant, but
simplified models exist, their observational testing is essential.
Moreover, such observational studies can provide an insight into
astrophysically important questions of driving turbulence and the
scales of energy injection.

The statistics of density, which is readily available in many cases,
is related in a non-trivial way with the magnetic and velocity fluctuations
(\cite{LG01}). Obtaining of the statistics of
velocity fluctuations from observations
is a problem that has been subjected to scrutiny recently (see \cite{Laz04}
and references therein). In what follows we shall discuss the problem
of determining the statistics of magnetic fields by observing fluctuation
of polarization arising from aligned dust and atoms.


\section{II. What does cause grain alignment?}

Starlight polarization measurements by Hall and Hiltner
 \cite{Hall, Hiltner} revealed that interstellar
dust grains are aligned. It did not take 
long  to realize that grains tend in most cases tend to
be  aligned with their long axes 
perpendicular to magnetic field. However, progress in theoretical 
understanding of the alignment
has been surprisingly slow in spite of the fact that great minds like
L. Spitzer and E. Purcell worked on grain alignment (see \cite{ST, P69, P75, P79, SM79}). 
The problem happened to be both
very and tough as a lot of relevant physics had to be uncovered. 

Currently, grain alignment theory is an exciting branch of 
research where a lot of insight
has been obtained lately (see reviews
\cite{Laz03, LC05}). We shall refer the reader to these
reviews which contain dozens of references to the original important work,
while here we present a few facts about dust alignment that are needed to
understand how well the alignment represents magnetic field direction. 

Originally it was widely believed that interstellar grains can be well aligned
by a paramagnetic or Davis-Greenstein mechanism \cite{DG}. This mechanism based
on the direct interaction of rotating grains with the interstellar 
magnetic field required magnetic fields that are stronger than those
obtained by other techniques.
Later, a pioneering work by Purcell \cite{P79} showed a way how to make grain 
alignment more
efficient. Purcell introduced several 
processes that
are bound to make grains very fast rotators and 
noticed that grains rotating at high rates are not 
so susceptible
to the randomization induced by gaseous collisions. For decades 
this became a standard
explanation for grain alignment, even though it could not explain several 
observational
facts, e.g. why  small grains are less aligned than the large ones 
(\cite{KM95}).

New physics of grain internal motion uncovered recently 
explains inefficiency of alignment of small grains by Purcell's mechanism. 
Lazarian \& Draine \cite{LD99a} found that small grains flip frequently 
due to the
coupling of rotational and vibrational degrees of freedom of a grain. 
As the result
regular torques, e.g. torques due to ejection of H$_2$ molecules, 
get averaged out
and grains rotate at thermal velocities. The paramagnetic alignment of 
thermally
rotating grains as we mentioned earlier is inefficient 
(see \cite{RL99}).
Interestingly enough, Lazarian \& Draine \cite{LD99b} found that coupling 
of rotational
and vibrational degrees of freedom happens most efficiently through 
a process the origin of which is related to {\it nuclear} spins!
This relaxation makes grains of size $10^{-5}{\rm cm}< a< 10^{-4}$~cm rotate 
thermally,
which makes the Purcell mechanism inefficient for most of dust in {\it diffuse}
 interstellar medium.

A group of alternative mechanisms of alignment related to the relative 
gas-grain
motion have their particular niches. The first mechanical
alignment mechanism was pioneered by Gold \cite{Gold}. Later work included 
driving grains
by ambipolar diffusion \cite{RH90} and Alfv\'en waves 
\cite{Laz94, Laz97, YL03}. Although
new efficient 
processes of mechanical alignment \cite{Laz95, LEO96}, this did not make
the mechanical alignment universally applicable. 
 
Quite recently, radiative torque (henceforth { RT}) have been accepted
as the most promising mechanism for explaining grain
alignment over vast expanses of interstellar space. Introduced first 
by Dolginov
\cite{Dog72} and Dolginov \& Mytrophanov \cite{DM76} the RT were
mostly forgotten till a more recent study by Draine \& Weingartner
\cite{DW96}, where their 
efficiency was demonstrated
using numerical simulations (see also \cite{DW97, WD03}, \cite{CL05b}).

The RT make use of interaction of radiation with a grain 
to spin the grain up. Unpolarized light can be presented as a superposition
of photons with left and right circular polarization. In general, the 
cross-sections of interaction of such photons with an irregular grain are
different. As the result of preferential extinction of photons with
a particular polarization the grain experiences regular
torques and gets spun up.

The predictions of RT mechanism are roughly 
consistent
with the molecular cloud extinction and emission polarimetry 
\cite{LGM}
and the polarization spectrum measured \cite{Hild}. 
RT have been
demonstrated to be efficient in a laboratory setup \cite{Abbas}. 
Evidence in favor of RT alignment was found for the data obtained at the interface of the
dense and diffuse gas (see \cite{Laz03}). 

While it was originally believed that RT cannot align grains
at optical depths larger than $A_v\approx 2$, Cho \& Lazarian
\cite{CL05b} demonstrated
that the efficiency of RT increases sharply with the grain size
and therefore bigger grains that exist within molecular clouds can be aligned
for $A_v$ more\footnote{Our studies reveal that for fractal
molecular clouds the alignment can be present for cores with $A_v$ of
30. In addition, as large grains do not flip frequently 
the Purcell alignment gets efficient as well.} than 10. 
Large grains may constitute an appreciable part
of the total mass of dust within a cloud, while still be marginal in terms
of light extinction. Therefore a non-detectable polarization in optical
and near infrared does not preclude substantial polarization to be present
in submillimeter. This makes submillimeter polarimetry the preferred tool
for studies magnetic fields and magnetic turbulence in molecular clouds.

To align grains with RT mechanism requires grains to be comparable to the
wavelength they interact with. Therefore grains smaller than 
$\sim 5\times 10^{-6}$~cm are not aligned by RT in typical interstellar
conditions\footnote{Such grains can still be aligned by strong UV flux
near some stars.}. These grains flip frequently and therefore they cannot
be aligned by \cite{P79} process according to \cite{LD99a}.
However, these grains can be aligned by the original Davis-Greenstein process
(see detailed calculations in \cite{Laz97, RL99}).
The degree of the expected alignment for typical interstellar conditions
does not exceed several 
 percent, but this alignment is sensitive to the {\it magnitude} of magnetic
field. The data in \cite{KM95} is consistent with the existence
of a population of small aligned grains. Such aligned grains may be revealed
with UV polarimetry.

\section{III. What does aligned dust reveal?}

Most of the processes produce grain alignment in respect to magnetic
field, even if the alignment mechanism is of non-magnetic nature.
This is true due to the rapid precession of grains about magnetic field.
This precession stems from the substantial magnetic moment that grains get
due to their rotation \cite{DM76}. The 
corresponding period of grain precession $\tau_L$ 
is $\sim 10^5 B_{3}^{-1} a_{-5}^2$~s, where the 
the external magnetic field is normalized over its typical interstellar value
of $3\times 10^{-6}$~G and grain size is chosen to be $a=10^{-5}$~cm.
 This means that for turbulent motions on time
scales longer than $\tau_L$ grain alignment in respect to magnetic 
field lines does not change as the consequence of the adiabatic invariant
conservation (see more discussion in \cite{Laz03}). 

If the alignment happens on the time scales shorter than $\tau_L$ the dust
orientation may not reflect the magnetic field. For the RT
such a fast alignment will happen with longer grain axes perpendicular 
{\it to the direction of radiation}, while the fast mechanical alignment will
happen with longer axes parallel to the flow\footnote{The rule of thumb for
mechanical alignment is that it tends to minimize the grain cross section
for the grain-flow interaction, while for RT is that
the grain precession is minimized.}. The mechanical and  RT
alignment takes place on the time scale of approximately\footnote{The mechanical alignment happens faster due to the fact that the flows are supersonic. This is an important difference to be considered for transient alignment, but
such an alignment is not typical for ISM.} gaseous damping
time, which is for interstellar medium is 
$\sim 10^{11} T^{-1/2}_{100}n_{-20}^{-1} a_{-5}$~s, where 
typical temperatures and densities of cold interstellar medium, which
are respectively 100K and $20$~cm$^{-3}$, were used for the normalization.
Note, that magnetic alignment takes place over even longer time scales,
namely, $\sim 10^{13} B_{3} a^{2}_{-5}$~s. Therefore in most cases the magnetic
field indeed acts as the alignment axis.

It is worth noting that the turbulent fluctuations with periods longer
than $\tau_L$ {\it do not} suppress alignment. The rapidly precessing
grains preserve their orientation to the local direction of magnetic field
and undergo the alignment even when this local direction is changing 
in space. In this respect grain alignment is a local process
that can reflect local direction of magnetic field for magnetic ripples
larger than $\tau_L V_A$, where $V_A$ is Alfv\'en speed. 
Whatever is the process
of alignment, if it aligns grains over timescale larger than $\tau_L$ the
alignment is {\it perpendicular or parallel} to magnetic field. 

The issues of the mechanism and the
efficiency of alignment is essential when averaging along the 
line of sight is involved. For instance, if we study turbulence in a
molecular cloud we would like to know whether the polarization signal that
we measure is coming from a thin skin layer of the cloud or from the whole
cloud volume. We also would like to know whether the direction
 of alignment changes along the line of sight.

It follows from the discussion above that in most cases
 the assumption that grains are aligned with long axes 
perpendicular to magnetic fields is a valid one\footnote{Mechanical alignment
by plasma streaming as well as Davis-Greenstein alignment when grains are
warmer than the ambient gas present exceptions from this rule.}.
 The fact, that RT have been identified
as a dominant alignment mechanism plus that this process seems to be efficient
even at substantial optical depths simplifies the interpretation of the 
polarization fluctuations in terms of magnetic field fluctuations.

If we limit ourselves to discussing grains aligned by RT, the grains
in diffuse interstellar medium can be considered perfectly aligned with
longer axes perpendicular to magnetic field. Optical and near infrared
emission from stars will be polarized via the differential extinction by
the aligned grains so that the direction of linear polarization will be
parallel to the local interstellar magnetic field. The variations of
magnetic field directions along the line of sight from the star to the 
observer will influence the resulting degree of polarization $p$ as well as
the positional angle of the polarization vector $\delta \phi$. These
observational parameters can be used to obtain the polarization statistics.

The geometry of lines of sight is also important for the interpretation of
interstellar polarization statistics.
 It can be easily shown (see \cite{LS90},
\cite{CL02b}) that when a signal is collected along
 the converging lines of sight (see Fig.~1), 
the turbulence provides a universal
$\theta^{-1}$ spectrum of fluctuations when the correlated lines of sight
are separated by an angle larger than the ratio of the
injection scale of turbulence and the distance to the stars, i.e. $L/d_{max}$.

\begin{figure}
\includegraphics[width=0.4\textwidth, height=0.20\textheight]{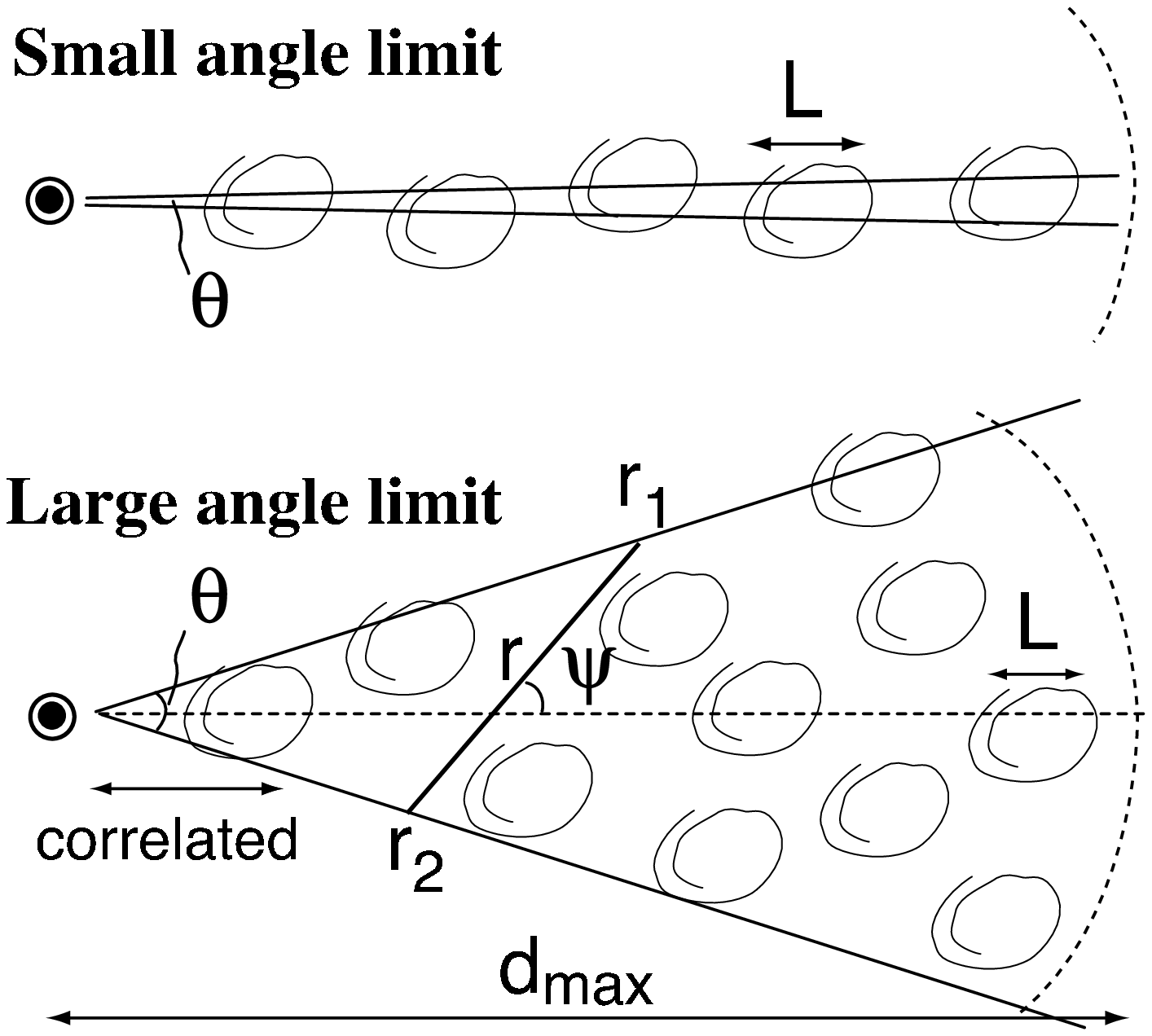}\hfil
\includegraphics[width=0.4\textwidth, height=0.18\textheight]{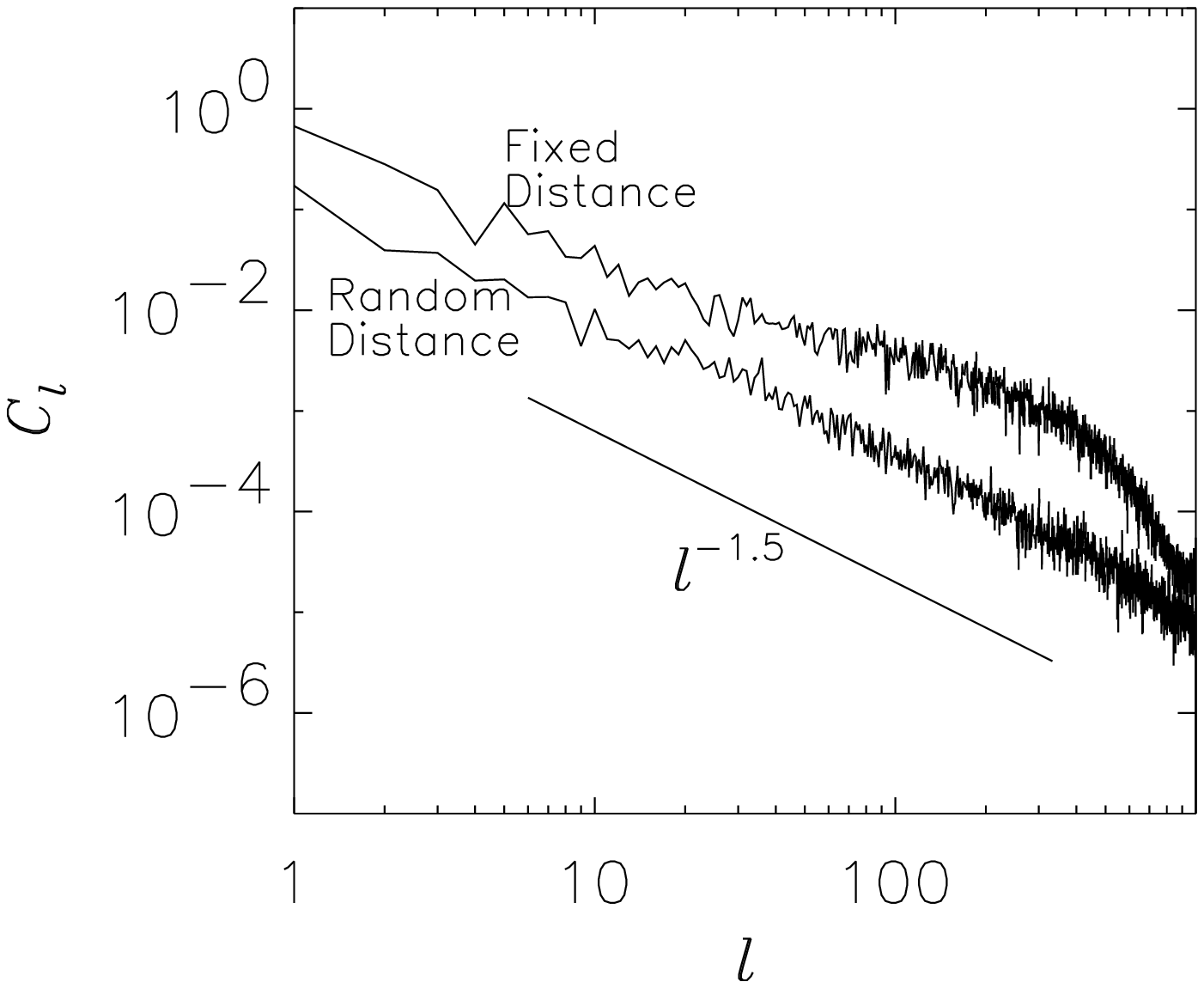}
\caption{{\it Left}: Two cases of measuring correlations with converging
lines of sight. For $\theta<L/d_{max}$ the underlying 
spectrum of magnetic turbulence can be
revealed. In the opposite limit the universal assymptotics $\theta^{-1}$
is present. {\it Right}: Spectra in terms of spherical harmonics for
the degree of polarization for 
stars at a fixed at at a random distance 
from the observer. From \cite{CL02b}.}
\end{figure}

The variations of $p$ were studied in \cite{Fos03} for a sample of
polarization measurement available in Heiles catalog. They provided a spectrum
$\theta^{-1.5}$ that was interpreted by \cite{CL02b} as the result
of the superposition of the contributions from distant and nearby stars 
in the presence of MHD turbulence in the interstellar dusty gas (see Fig.~1). 
A more systematic study of polarization using only close and only distant
stars could test the predictions in \cite{CL02b} . It shall also
be able to establish the correlation scale of galactic turbulence $L$.
Interestingly enough, confirming that the polarization fluctuations in Milky
Way arise from MHD turbulence is important from the point of view of CMB 
studies. Aligned dust provides an important contribution to polarized microwave
foreground (see \cite{LF03} and references therein). This
foreground must be removed to reveal the polarized 
emission originated in the Early Universe. MHD turbulence has a well defined
statistics and therefore this open ways for more reliable ways of weeding out
such a foreground.

\section{IV. What is atomic alignment?}

 As we discussed above,
polarimetry of aligned dust provides one of the ways of studying 
MHD turbulence. Polarimetry of some molecular lines using Goldreich-Kylafis
effect \cite{GK}
 have been shown recently to be a good tool for magnetic field  studies 
in molecular clouds (\cite{Girart}). 
Here we 
discuss yet another promising technique to study magnetic fields that
employs
optical and UV polarimetry. This technique can be used for 
interstellar\footnote{Here interstellar is understood in a 
general sense, which, for instance
includes relection nebulae.}
and circumstellar studies as well as for studies of magnetic fields in
QSOs and other exotic objects.  
The technique is based on
the ability of atoms to be aligned by external radiation in their ground state
and be realigned through precession in magnetic  field.

It has been known that  
atoms can be aligned through the interactions with the anisotropic flux
of resonance emission (see review \cite{Happer}). Alignment is  understood here in terms of orientation
of vector angular momentum
$\bf J$, if we use the language of classical mechanics. In quantum
terms  this means a difference in the population of sublevels corresponding to
the projections of angular momentum to the quantization axis. We will
call atomic alignment as the alignment of $\bf J$ in the ground state.
This state is long lived and therefore is being affected by
weak magnetic field.

Atomic alignment has been studied in laboratory in relation with early 
day maser 
research (\cite{Hawkins}). Alignment of atoms in ground state can change
the optical properties of the medium.  This effect was noticed and made
use of for the interstellar case 
by \cite{Varsha} in case of hyperfine structure of the ground state, for 
fine structure of the ground state by, e.g. \cite{Lee}. Here we shall be
concerned with the variations of degree of polarization $p$ and the positional
angle $\delta \phi$ that is caused by atomic alignment.
  
A possibility of using atomic alignment to study weak 
magnetic field in diffuse media was first discussed in \cite{Landi} 
for ideal two-level atoms with fine 
structures and polarization of emissions from these atoms  were 
considered for a very restricted geometry of observations,
namely, when magnetic field is aligned with 
line of sight and perpendicular to the incident light. A discussion
of alignment of various atoms and ions for arbitrary orientations of
magnetic fields towards the line of sight and the direction to the
illumination source is given in \cite{YL05}.

Similar to the case of interstellar dust, the rapid precession of atoms
in magnetic field makes the direction of polarization sensitive to the
direction of underlying magnetic field. As the precession of magnetic moments
of atoms is much faster than the precession of magnetic moments of grains 
atoms can reflect much more rapid variations of magnetic field. More 
importantly, alignable atoms and ions (see Table~1) 
can reflect magnetic fields in the
environments where either properties of dust change or the dust cannot survive.
This
opens wide avenues for the magnetic field research in circumstellar regions,
interstellar medium, interplanetary medium, intracluster medium and quasars 
etc. 
In addition, the degree of alignment $p$ can help in determining
 the angle between the direction towards the
illumination source and the direction of magnetic
field. This information is not available by any other technique.

\begin{table}
\begin{tabular}{||c|c|c|c|c|c|c||}

\hline
 Atom&
 Nuclear spin&
 Lower state&
 Upper state&
 Wavl(\AA)&
 Pol(emi)&
 Pol(abs)\tabularnewline
\hline 
\hline 
{Na I}&
$\frac{3}{2}$&
$1S_{1/2}$&
$2P_{3/2}$&
5891.6&
Y&
Y\\
\cline{4-7}
& &  &
$2P_{1/2}$&
5897.6&
Y&
Y\\
 \hline
{N V}&
$1$&
  $1S_{1/2}$&
$2P_{3/2}$&
1238.8&
Y&Y\\
\cline{4-7}
& & &
$2P_{1/2}$&1242.8&Y&Y\\
\hline
 Al III&
$\frac{5}{2}$&
$1S_{1/2}$&
$2P_{3/2}$&
 1854.7&M&M\\
\cline{4-7}
&& &
$2P_{1/2}$ & 1862.7&M&M\\
\hline
 H I&
$\frac{1}{2}$&
$1S_{1/2}$&
$2P_{1/2,3/2}$&
912-1216&Y&N\\
\hline
 N I&
1&
$4S^o_{3/2}$&
$4P_{1/2,3/2,5/2}$&
$865-1201$&Y&Y\\
\cline{1-2} \cline{5-5}
 O II&
0& & &
 375-834& & \\
\hline
 O I&
0& 
$3P_2$&
$3S_1$&
911-1302.2&M&Y\\
\hline
Cr II&0 &$6S_{5/2}$ & 
$6P^o_{3/2,5/2,7/2}$&
2056, 2062, 2066&Y
&Y \\
\hline
 C II&
0&
$2P^o_{3/2}$&
$2P_{3/2}$&
904.1&Y&Y\\
\cline{4-5}
& & &
$2D_{3/2,5/2}$&
1335.7& & \\
\hline
O IV&
0&
$2P^o_{3/2}$&
$2P_{3/2}$&
554.5&Y&Y\\
\cline{4-5}
& & &
$2D_{3/2,5/2}$&
239, 790& & \\
\hline
 C I&
0&
$3P_{1,2}$&
$3P^o_{0,1,2}$&
1118-1657&Y&Y\tabularnewline
\cline{4-5}
 &
 &
 &
$3D^o_{1,2,3}$&
 1115-1561& & \\
\hline
O III&0 &$3P_{0,1,2}$ & 
$3P^o_{0,1,2}$&
304, 374, 703&Y
&Y \\
\cline{4-5}
& & & $3D^o_{1,2,3}$&267, 306, 834& &\tabularnewline
 \hline
\end{tabular}
\caption{Note: only polarizable and alignable components are listed. "M" stands for marginal polarization. From \cite{YL05}}
\end{table}

\section{V. How do atoms get aligned?}

The basic idea of the atomic alignment is simple. The alignment is caused by
the anisotropic deposition of angular momentum from photons. In typical
 astrophysical situations the radiation
flux is anisotropic. As the photon
spin is along the direction of its propagation, we expect that atoms
that scattered the radiation can be  aligned
in terms of their angular momentum. Such an alignment happens in terms of 
the projection of angular momentum
to the direction of the incoming light. It  is clear that to have the
alignment of the ground state, the atom should have non-zero angular
momentum in its ground state. Therefore fine or hyperfine structure
of the ground state is necessary to enable various projection of
angular momentum to exist in the ground state. Whether polarization arising
from aligned atoms is
parallel or perpendicular to magnetic field depends on the angle that
the anisotropic radiation makes with the magnetic field.

Let us discuss a toy model that provides an intuitive  insight into  the 
physics of atomic alignment.
First of all, consider a toy model of an
atom with lower state corresponding to the total angular momentum
$I=1$ and the upper state corresponding to angular momentum $I=0$.
If the projection of the angular momentum to the direction of the
incident resonance photon beam is $M$, for the lower state $M$ can
be $-1$, $0$, and $1$ , while for the upper state $M=0$ (see Fig.~2{\it left}). 
The unpolarized
beam contains an equal number of left and right circularly polarized
photons which projection on the beam direction are 1 and -1. Thus
absorption of the photons will induce transitions from $M=-1$ and
$M=1$ states. However, the decay of the upper state happens to all
three levels. As the result the atoms get accumulated at the $M=0$ ground
state from which no excitations are possible. As a result of that the optical
properties of the media (e.g. absorption) would change\footnote{
Not every type of alignment affects  the polarization of the
scattered  of absorbed  radiation. Interestingly
enough,  alignment that we discuss within the toy model
does not affect
polarization of resonant light. 
To have emission polarized, the alignment on the ground state should be transfered to the excited level. Therefore atoms with more complex structure
of the excited levels should be considered.}.

This above toy model can also exemplify the
 role of collisions and magnetic field. Without collisions one may expect 
that all atoms
reside eventually at the state of $M=0$. Collisions, however, redistribute
atoms with different states. However, as disalignment of the ground
state requires spin flips, it is less efficient than one can naively
imagine. Calculations by \cite{Hawkins} show that to disalign sodium
one requires more than 10 collisions with paramagnetic atoms and experimental
data by \cite{Kastler} support this. This reduced sensitivity of aligned
atoms to disorienting collisions makes the effect important for various
astrophysical environments.

\begin{figure}
\includegraphics[width=0.32\textwidth, height=0.18\textheight]{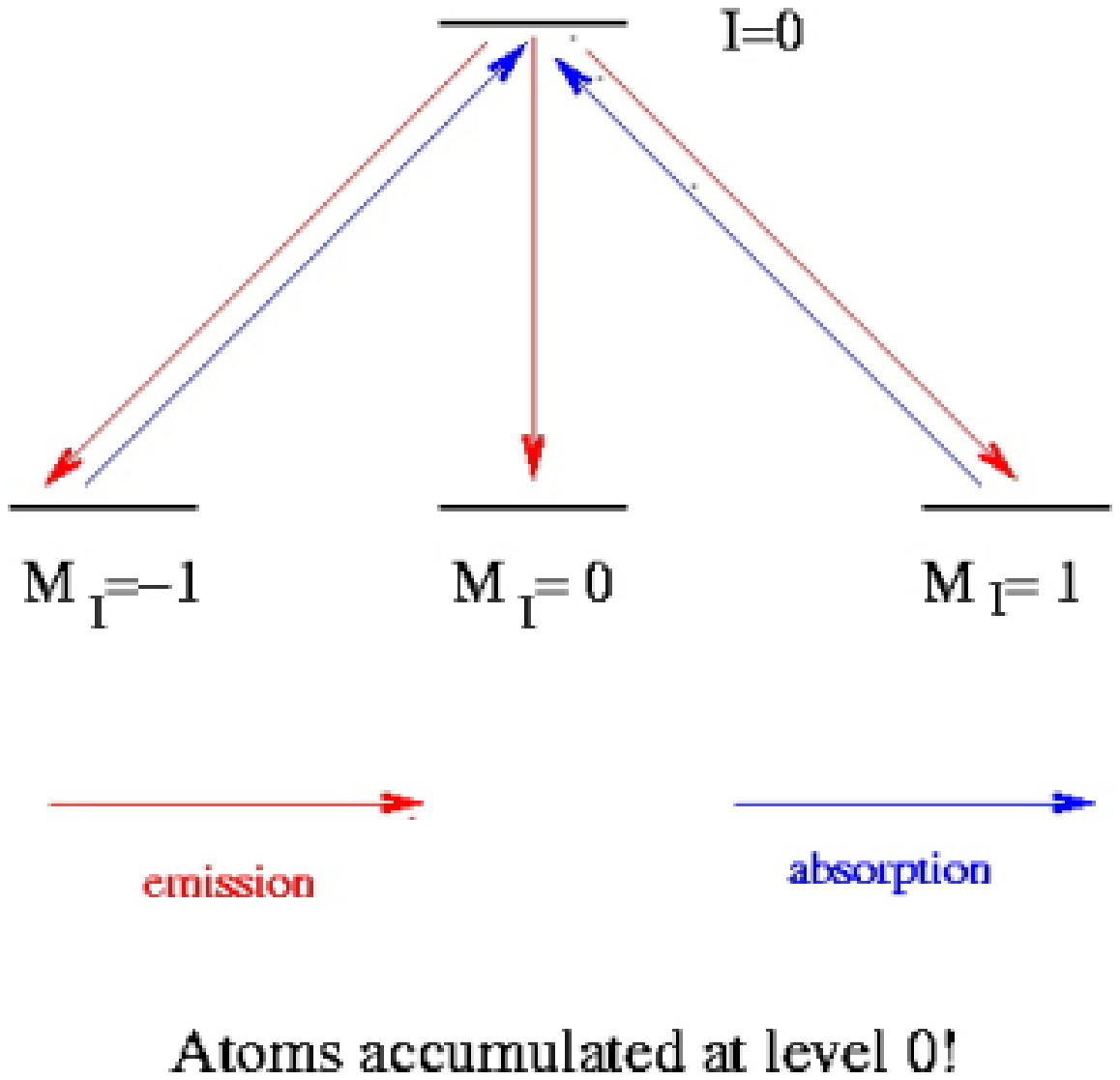}\hfil
\includegraphics[width=.32\textwidth, height=0.17\textheight]{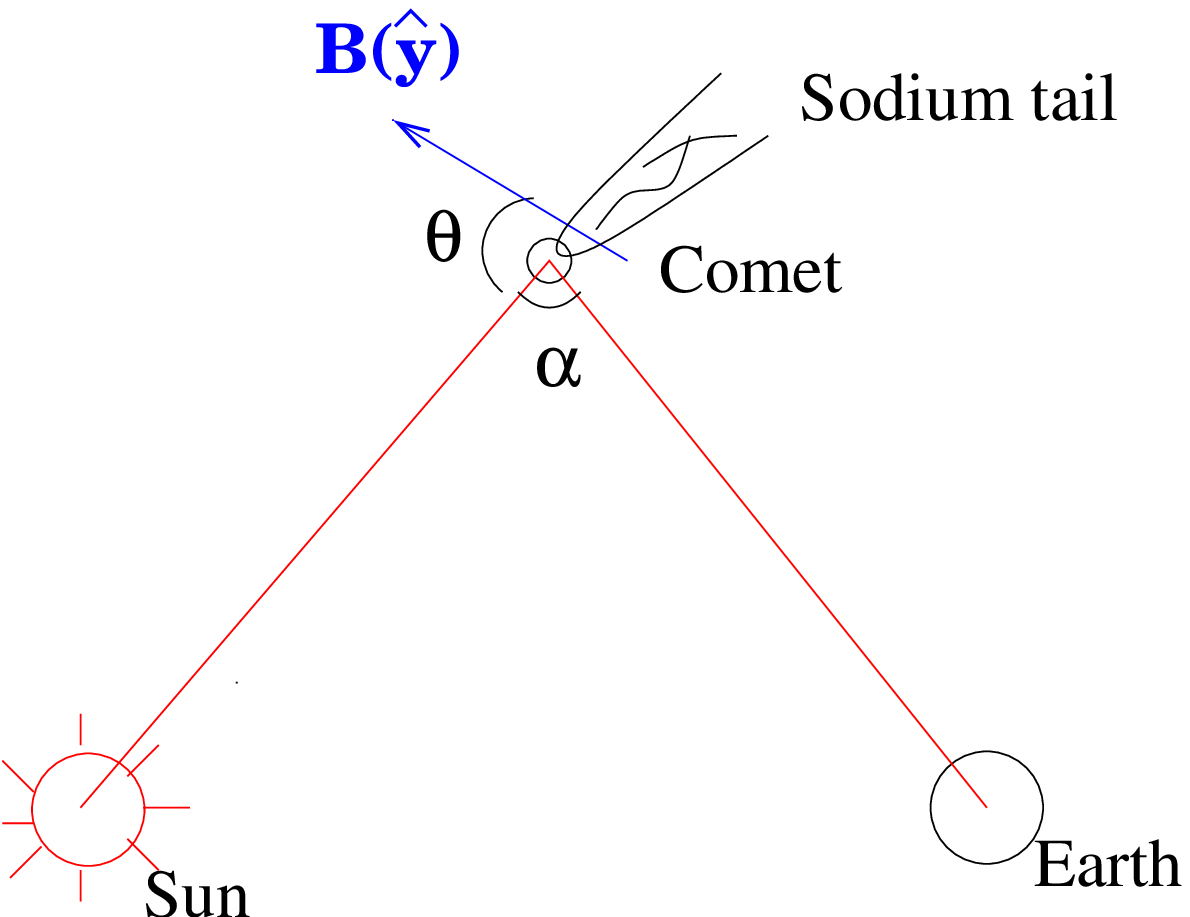}\hfil
\includegraphics[width=.31\textwidth, height=0.18\textheight]{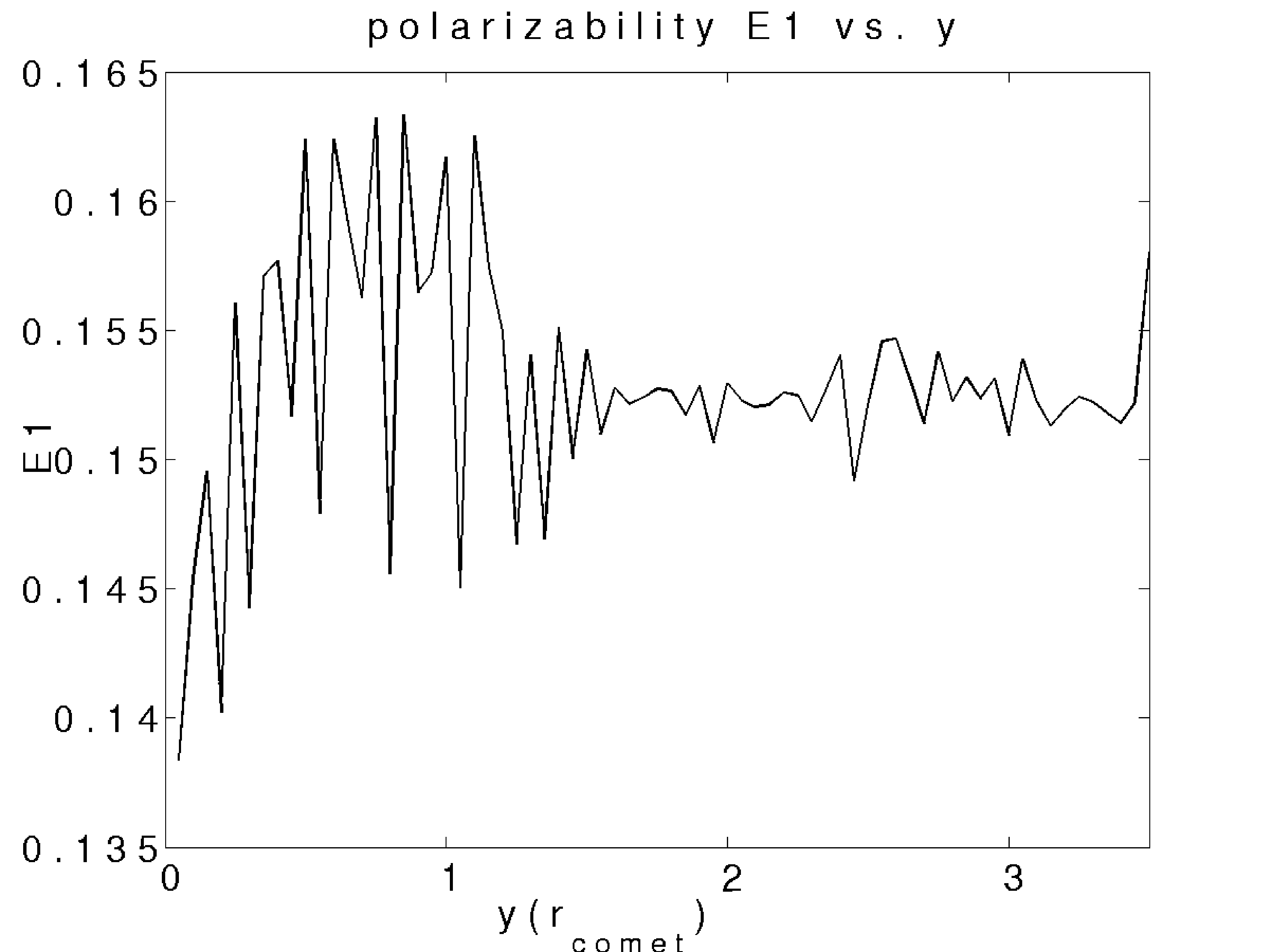}
\caption{{\it Left}: A toy model to illustrate how atoms are aligned by anisotropic light.
Atoms get accumulated in the ground state $M_I=0$, as radiation removes atoms from the ground states $M_I=1$ and $M_I=-1$; {\it middle}: Resonance scattering of solar light by sodium tail from comet; 
{\it right:} Polarization caused by sodium aligned in the comet wake. Spatial and 
temporal fluctuations of polarization carry the information on MHD turbulence.}
\end{figure}

Magnetic field would also mixes up different $M$ states. However, it
is clear that the randomization in this situation will not be complete
and the residual alignment would reflect the magnetic field direction
in respect to the observer. Magnetic mixing happens if the angular
momentum precession rate  is higher than the rate of the
excitation of atoms from the ground state, which is true
for many astrophysical conditions. 

Note that in order to be aligned, first the atoms should have enough 
degree of freedom, namely, the quantum number of angular momentum must be 
$\ge 1$. Second, incident flux must be anisotropic. 
Moreover, the collisional rate should not be too high. While the latter
requires special laboratory conditions, it is the
case for many astrophysical environments such as the outer
layers of stellar atmosphere, interplanetary,
interstellar, and intergalactic media, etc.

As long as these conditions are satisfied, atoms can be aligned within either 
fine or hyperfine structure. For light elements,
the hyperfine splitting is very small and the line components overlap
to a large extent. However, for resonant lines, the hyperfine interactions 
causes 
substantial 
precession of electron angular momentum ${\bf J}$ about
the total angular momentum ${\bf F}$ before spontaneous emission. Therefore 
total angular momentum should be considered and the $FM_{F}$
basis must be adopted (\cite{Walkup}). For alkali-like atoms,
 hyperfine structure should
be invoked to allow enough degrees of freedom to harbor the alignment and to 
induce the corresponding polarizations.

In terms of time scales, we have a number of those, which makes the problem
interesting and allows getting additional information about environments.
The corresponding rates 
are 1) the rate of the precession $\tau_L^{-1}$, 2) the rate of the photon
arrival $\tau_A^{-1}$, 3) the rate of collisional randomization $\tau_R^{-1}$,
4) the rate of the transition within fine or hyperfine structure $\tau^{-1}_T$.
In many case
$\tau_L^{-1}>\tau_A^{-1}>\tau_R^{-1}>\tau_T^{-1}$. 
Other relations are possible, however.
If the Larmor precession gets comparable with any of the other rates,
it is possible to get information about the {\it magnitude} of magnetic
field. Another limitation of our approach is that we consider that $\tau_L^{-1}$ is much smaller that the rate of the decay of the excited state, which
means that we disregard the Hanle effect.

\section{VI. How to study interplanetary turbulence without space missions?}

As an illustration, we discuss here a synthetic observation of a comet wake. 
Though the abundance of sodium in comets is very low, its high efficency of
 scattering Sun light makes it a good tracer 
(\cite{Thomas, Cremon}). The sodium atoms ejected from the comets are 
aligned by the solar radiation. Distant from comets, the Sun can be 
considered a point source. As shown in Fig.~2{\it Middle}, the geometry of the 
scattering is well defined, i.e., the scattering angle $\alpha$ is known. 
The polarization of the sodium emission thus provides an exclusive 
information of the magnetic field in the comet wake.  Embedded in Solar 
wind, the magnetic field is turbulent in a comet wake. We take a 
data cube directly from MHD simulations of a comet wake. Depending on its 
direction, the embedded magnetic field alter the degree of 
alignment and therefore polarization of the light scattered by the aligned 
atoms. Therefore, fluctuations in the linear polarization are expected from 
such a turbulent field (see Fig.~2{\it right}). For interplanetary studies,
one can investigate not only spatial, but also temporal variations
of magnetic fields. This can allow cost effective way of studying
interplanetary magnetic turbulence at different scales.

\section{VII. What can we learn about MHD turbulence?}

Astrophysical environments provide unique opportunities for studying MHD
turbulence  
at Reynolds and magnetic Reynolds numbers that are not available
through either numerical simulations or laboratory experiments. This allows
to test our theoretical constructions to see whether numerics is adequate.
 
It is difficult to overestimate the importance of understanding better 
astrophysical turbulence including its spectra, injection and dissipation
scales. Magnetic field statistics is 
complementary to the velocity statistics that can be studied using 
Doppler shifted spectral lines (see \cite{Laz94}).

Aligned dust and aligned atoms provide a unique way of studying
magnetic turbulence in various astrophysical environments from molecular
clouds to quasars.  Recent progress
in theories of atomic and grain alignment gives us confidence in interpreting
polarization in terms of underlying magnetic fields.

Aligned atoms provide a possibility of
 studying temporal variations arising
from interplanetary turbulence.
 Techniques that use aligned atoms and aligned grains are synergetic
for studies of magnetic turbulence.

{\bf ACKNOWLEDGEMENTS}\\
We are grateful to Jungyeon Cho for  
supplying us with MHD simulations of a comet wake. The work was supported 
by the NSF grant  AST 0243156.

\end{document}